\documentclass[12pt]{iopart}
\usepackage{listings}
\usepackage{hyperref}
\usepackage{graphicx}
\usepackage[utf8]{inputenc}
\usepackage{tikz}
\usepackage[labelfont=bf]{caption}

\begin{document}

\title{Speeding HEP Analysis with ROOT Bulk I/O}

\author{B Bockelman$^1$, Z Zhang$^2$ and O Shadura$^2$}
\address{$^1$ Morgridge Institute for Research, Madison, WI 53715, USA}
\address{$^2$ Holland Computer Center, University Nebraska $-$ Lincoln, Lincoln, NE 68588, USA}
\ead{bbockelman@morgridge.org}
\vspace{10pt}

\begin{abstract}
Distinct HEP workflows have distinct I/O needs; while ROOT I/O excels at serializing complex C++ objects common to reconstruction, analysis workflows typically have simpler objects and can sustain higher event rates. To meet these workflows, we have developed a “bulk I/O” interface, allowing multiple events’ data to be returned per library call. This reduces ROOT-related overheads and increases event rates -- orders-of-magnitude improvements are shown in microbenchmarks.

Unfortunately, this bulk interface is difficult to use as it requires users to identify when it is applicable and they still “think” in terms of events, not arrays of data. We have integrated the bulk I/O interface into the new RDataFrame analysis framework inside ROOT. As RDataFrame’s interface can provide improved type information, the framework itself can determine what data is readable via the bulk IO and automatically switch between interfaces. We demonstrate how this can improve event rates when reading analysis data formats, such as CMS’s NanoAOD.
\end{abstract}

\section{Introduction}
\label{sec:introduction}

LHC experiment event data models are very complex and slow to read. The problem is that experiments do not care because input I/O time is minimal compared to the reconstruction process. Another critical factor is that experiments care about volume because they have lots of expensive disks.

For analysis case, the situation is different, since the data model is often more straightforward. The same case is about data volume used during the analysis phase, and there will be generated smaller a data volume and often used from SSD (NVMe). It causes minimal CPU costs and allows to iterate over events many times quickly. 

ROOT IO is an incredibly flexible format. It can easily store the complex objects that correspond to the experiment’s data. In the same time, ROOT has high overheads for the serialization of simple objects.


\section{Bulk IO}
\label{sec:bulkio}

The typical mechanism for iterating through data in a \texttt{TTree} is a handwritten for-loop. ROOT uses a API shown in Listing \ref{getentry} to read objects from a branch (TTree is a structure that contains one or multiple TBranches). This function runs in two steps. First, it searches the underlying storage medium for the basket where the event is located and then read the basket into a memory buffer. The TBasket is the data structure that represents the in-memory buffer. ROOT decompresses the buffer and put the uncompressed buffer in so-called ``kernel" space. In the second step, once the basket appears in memory,  GetEntry deserializes the requested event from the kernel-space buffer and copy it to user-space buffer.

\lstset{
  basicstyle=\ttfamily,
  frame=single,
  xleftmargin=.15\textwidth, xrightmargin=.15\textwidth
}
\begin{lstlisting}[label={getentry}]
  Int_t TBranch::GetEntry(Long64_t entry)
\end{lstlisting}
\vspace{-10pt}
\captionof{lstlisting}{GetEntry in TBranch}

When user application is computationally expensive, the cost of library calls, frequently deserializing objects and copying data between memory buffers are amortized to effectively nothing. To overcome such overheads, we introduce a new interface for ROOT to copy all events in an on-disk \texttt{TBasket} directly to a user-provided memory buffer. For the simplest cases - primitives and C-style arrays of primitives, the serialization can be done without a separate buffer or ``fixing up" pointer contents. The user can request the serialized data or deserialized data to be delivered to the user buffer. By requesting the serialized data directly and deserializing directly in the event loop, the user can avoid an expensive scan from main memory.

Pragmatically, the user will not implement code for deserializing data themselves: rather, we have provided a header-only C++ facade around the data, allowing the user to work with a proxy object.  This allows the compiler to inline the deserialization code in the correct place.







\section{Implementation}
\label{sec:implementation}

Bulk IO interface is a set of APIs that are built in the existing ROOT IO framework. The user can choose between regular APIs and Bulk IO APIs. We implement Bulk IO in three common use cases: \textit{TBranch}, \textit{TTreeReader} and \textit{RDataFrame}. We discuss about our interface design and integration in this section.

\subsection{Bulk IO in TBranch}

Listing \ref{bulkapi} shows Bulk IO API in TBranch in which two input arguments need to be parsed into the function: \textit{entry} and \textit{user\_buf}. the entry defines an event index number indicating which event the function is going to read. The \textit{user\_buf} parses an user-space TBuffer structure as a reference into the function. In the end of the function call, the \textit{user\_buf} should contain the whole basket of data that contains the inquiry event.

\lstset{
  basicstyle=\footnotesize\ttfamily, frame=single,
  xleftmargin=.05\textwidth, xrightmargin=.05\textwidth
}
\begin{lstlisting}[label={bulkapi}]
Int_t TBranch::GetBulkEntries(Long64_t entry, TBuffer &user_buf)
\end{lstlisting}
\vspace{-10pt}
\captionof{lstlisting}{Bulk API in TBranch}


It is worthwhile to mention that \textit{GetBulkEntries} deserializes events on-the-fly when the data read into the \textit{user\_buf}. Thus no further manipulation is required for user applications. An user can later on access an event in the basket using the code snippet shown in Listing \ref{loop} where \textit{T} is the object type and \textit{idx} is the event index in the \textit{user\_buf}.

\lstset{
  basicstyle=\ttfamily, frame=single,
  xleftmargin=.08\textwidth, xrightmargin=.08\textwidth
}
\begin{lstlisting}[label={loop}]
 *reinterpret_cast<T*>(user_buf.GetCurrent())[idx]
\end{lstlisting}
\vspace{-10pt}
\captionof{lstlisting}{Loop over user\_buf}

\subsection{Bulk IO in TTreeReader}

\lstset{
  language=C++,
  keywordstyle=\color{blue},
  stringstyle=\color{red},
  commentstyle=\color{green},
  basicstyle=\ttfamily, frame=single,
  xleftmargin=.09\textwidth, xrightmargin=.09\textwidth
}
\begin{lstlisting}[label={ttreereader}]
TTreeReader myReader("T", hfile);
TTreeReaderValue<float> myF(myReader, "myFloat");
Long64_t idx = 0; 
Float_t sum = 1; 
while (myReader.Next()) { 
   sum += *myF;
}
\end{lstlisting}
\vspace{-10pt}
\captionof{lstlisting}{Access to Events using TTreeReader}


\textit{TTreeReader} is an interface for an user to access simple object (primitives, arrays, etc.) in a ROOT file. \textit{TTreeReaderValue} is the interface to access primitives and \textit{TTreeReaderArray} is the interface to access arrays (each event is either an fixed-size or varialbe-size array). Listing \ref{ttreereader} shows a code sample that uses \textit{TTreeReaderValue} to read events (floats) from a file. \textit{TTreeReader} internally relies on \textit{GetEntry} to access events.

\lstset{
  basicstyle=\footnotesize\ttfamily, frame=single,
  xleftmargin=0\textwidth,
  xrightmargin=0\textwidth
}
\begin{lstlisting}[label={serialized}]
Int_t TBranch::GetEntriesSerialized(Long64_t entry, TBuffer &user_buf)
\end{lstlisting}
\vspace{-10pt}
\captionof{lstlisting}{Bulk API in TTreeReaderFast}


We design a Bulk API - \textit{GetEntriesSerialized} in TBranch shown in Listing \ref{serialized}. We introduce a new interface - TTreeReaderFast that uses \textit{GetEntriesSerialized} to function as TTreeReader. Unlike \textit{GetBulkEntries}, GetEntriesSerialized does not deserialize events while reading basket into \textit{user\_buf}. Instead, it waits until the user calls \textit{*myF}. Dereference operator invokes the appropriate deserialization code. 

\subsection{Bulk IO in RDataFrame}

During our work, Bulk IO is also integrated into RDataFrame \cite{rdataframe} which is a python Pandas \cite{pandas} like data analysis framework for ROOT users. RDataFrame provides a proxy interface - RDataSource (RDS). It allows RDF to read arbitrary data formats such as TTree, CSV, etc..

\lstset{
  basicstyle=\ttfamily,
  breaklines=true,
  xleftmargin=0.1\textwidth,
  xrightmargin=0.1\textwidth
}
\begin{lstlisting}[label={serializedcount}]
Int_t TBranch::GetEntriesSerialized(Long64_t entry, TBuffer &user_buf, TBuffer *count_buf)
\end{lstlisting}
\vspace{-10pt}
\captionof{lstlisting}{Bulk API in RDataFrame}


We define a new Bulk API \textit{GetEntriesSerialized} shown in Listing \ref{serializedcount}. The only difference from Listing \ref{serialized} is that there is one more argument \textit{count\_buf}. Listing \ref{serialized} actually calls Listing \ref{serializedcount} and set the \textit{count\_buf} as nullptr. \textit{count\_buf} is used to store array length information when events in RDataFrame are arrays. Variable-size arrays need such information to deserialize the butter into multiple individual arrays.
\section{Evaluation}
\label{sec:evaluation}

\subsection{Experiments}

All tests are conducted on a desktop Intel i5 4-Core @ 3.2GHz. A TTree with 100 million float values is read with different APIs. We tested three different use cases: GetBulkEntries, TTreeReaderFast and RDataSource.

\subsection{Results}

\begin{figure}[!ht]
\centering
\includegraphics[height=2.5in, width=3.5in]{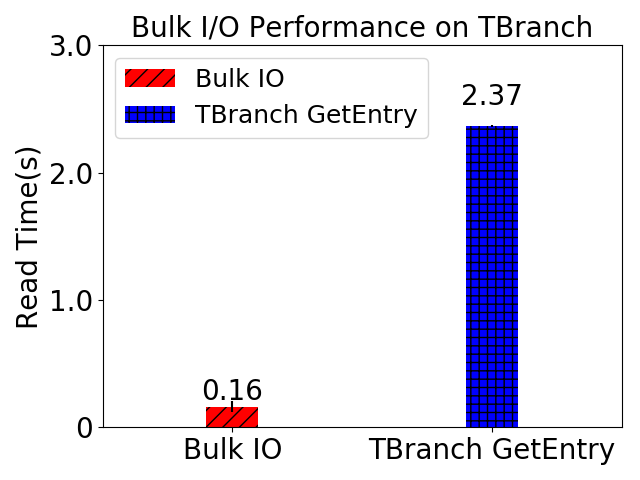}
\vspace*{-2mm}
\caption{Performance between \textit{GetEntry} and \textit{GetBulkEntries}.}
\label{fig:tbranch}
\end{figure}

\begin{figure}[!ht]
\centering
\includegraphics[height=2.5in, width=3.5in]{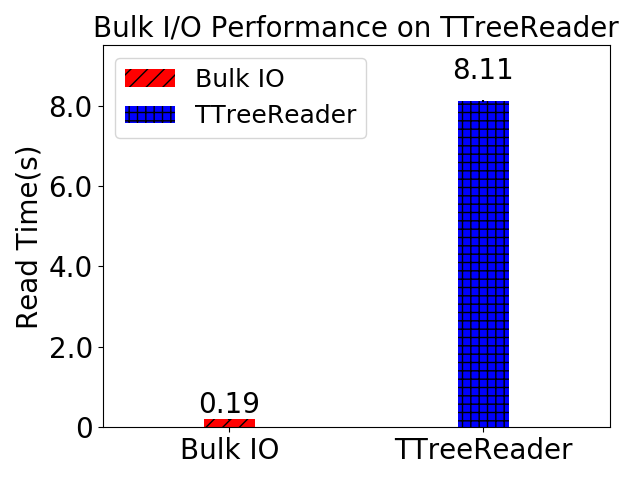}
\vspace*{-2mm}
\caption{Performance between \textit{TTreeReader} and \textit{TTreeReaderFast}.}
\label{fig:ttreereader}
\end{figure}

Figure \ref{fig:tbranch} shows the time spent on iterating all events in the \textit{TTree} with \textit{GetEntry} and \textit{GetBulkEntries}. Figure \ref{fig:ttreereader} shows the read time between \textit{TTreeReader} and \textit{TTreeReaderFast}. As shown in the figures, Bulk IO spends 10+ times less than \textit{GetEntry} and \textit{TTreeReader}. Bulk IO in both use cases spends similar time on reading events. \textit{TTreeReader} interface spends more than 3 times reading events than \textit{GetEntry} due to the overheads of TTreeReader itself (TTreeReader internally calls \textit{GetEntry}). 

\begin{figure}[!ht]
\centering
\includegraphics[height=2.5in, width=3.5in]{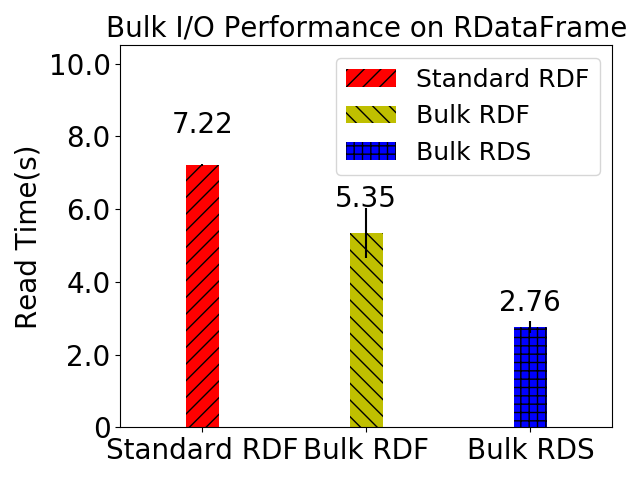}
\vspace*{-2mm}
\caption{Performance improvements on RDataFrame with Bulk IO.}
\label{fig:rdataframe}
\end{figure}

Figure \ref{fig:rdataframe} shows the results of Bulk IO in RDataFrame. In the figure, the standard RDF shows the performance by using regular RDataFrame function calls. Bulk RDF and Bulk RDS show the result of Bulk APIs. The difference is that, Bulk RDS test detaches RDataSource from RDataFrame stack and run the test directly through RDS function calls. As shown in Figure \ref{fig:rdataframe}, Bulk RDS outperforms standard RDF by more than 2 times. In addition, RDataFrame has extra overheads compared to RDataSource (RDataFrame internally relies RDataSource), therefore Bulk RDF runs slower than Bulk RDS, but still outperforms standard RDF.


\section*{Acknowledgments}

This work was supported by the National Science Foundation under
Grant ACI-1450323. This research was done using resources provided
by the Holland Computing Center of the University of Nebraska.


\section*{References}

\begin{thebibliography}{10}

\bibitem{root}
Brun R and Rademakers F ``ROOT - An object oriented data analysis framework", \textit{Nucl. Instr. Meth. Phys. Res.} \textbf{389} (1997) 81-86
\bibitem{rdataframe}
Guiraud E, Naumann A and Piparo D ``RDataFrame: functional chains for ROOT data analyses", (2017) doi: 10.5281/zenodo.260230. url: https://doi.org/10.5281/zenodo.260230.
\bibitem{bulk-io}
Bockelman B, Zhang z and Pivarski J ``Optimizing ROOT IO For Analysis", \textit{J. Phys.: Conf. Ser.}, \textbf{1085} (2018) 032012
\bibitem{pandas}
Mckinney W ``pandas: a Foundational Python Library for Data Analysis and Statistics", \textit{PyHPC 2011 : Python for High Performance and Scientific Computing}, (2011)
\end{thebibliography}
\end{document}